\documentclass[useAMS,usenatbib]{mn2e}
\usepackage{natbib}
\usepackage{graphicx}

\input{psfig}
\def\ga{\mathrel{\mathchoice {\vcenter{\offinterlineskip\halign{\hfil
$\displaystyle##$\hfil\cr>\cr\sim\cr}}}
{\vcenter{\offinterlineskip\halign{\hfil$\textstyle##$\hfil\cr>\cr\sim\cr}}}
{\vcenter{\offinterlineskip\halign{\hfil$\scriptstyle##$\hfil\cr>\cr\sim\cr}}}
{\vcenter{\offinterlineskip\halign{\hfil$\scriptscriptstyle##$\hfil
\cr>\cr\sim\cr}}}}}
\title[Extreme value statistics of smooth random Gaussian fields]{Extreme value statistics of smooth random Gaussian fields}

\author[S. Colombi, O. Davis, J. Devriendt, S. Prunet \& J. Silk]{St\'ephane
  Colombi$^1$\thanks{E-mail: colombi@iap.fr}, Olaf Davis$^2$, Julien
  Devriendt$^2$, Simon Prunet$^1$  \& Joe Silk$^{1,2}$\\
$^1$Institut d'Astrophysique de Paris, CNRS UMR 7095 and UPMC, 98bis, bd Arago, F-75014 Paris, France\\
$^2$Astrophysics, University of Oxford, Keble Road, Oxford OX1 3RH, UK}
\begin{document}
\voffset -1cm
\date{\today}
\pagerange{\pageref{firstpage}--\pageref{lastpage}} \pubyear{2010}
\maketitle
\label{firstpage}
\begin{abstract}
We consider the Gumbel or extreme value statistics 
describing the distribution function 
$p_{\rm G}(\nu_{\rm max})$ of the maximum values of a random 
field $\nu$ within patches of fixed
size. We present, for smooth Gaussian random fields in two and 
three dimensions, an analytical estimate of $p_{\rm G}$ 
which is expected to hold in a regime where local maxima 
of the field are moderately high and weakly clustered.

When the patch
size becomes sufficiently large, the negative of the logarithm of the cumulative 
extreme value distribution is simply equal to the average of the Euler Characteristic
of the field in the excursion $\nu \ge \nu_{\rm max}$ inside the
patches. The Gumbel statistics therefore represents an interesting alternative probe
of the genus as a test of non Gaussianity, e.g. in cosmic
microwave background temperature maps or in three-dimensional galaxy
catalogs. It can be approximated, except in the remote positive tail, by a negative
Weibull type form, converging slowly to the expected Gumbel type
form for infinitely large patch size. Convergence is facilitated
when large scale correlations are weaker.

We compare the analytic predictions to numerical experiments for the case
of a scale-free Gaussian field in two dimensions, achieving impressive
agreement between approximate theory and measurements. 
We also discuss the generalization of our formalism to non-Gaussian
fields.
 
\end{abstract}

\section{Introduction}
Gumbel or extreme value statistics are concerned with the extrema of samples
drawn from random distributions \citep{Gumbel}. 
In the case of sample means, the Central Limit Theorem
states that the means of many samples of size $N$ drawn from some distribution will
be normally distributed in the large-$N$ limit; analogously, it can be shown that in the
same limit the cumulative distribution of the sample maximum or minimum $\nu$
will tend to one of the family of the following functions,
\begin{equation}
G_{\rm \gamma_{\rm G}}(\nu)=\exp\left[ -\left( 1+\gamma_{\rm G} y \right)^{-1/\gamma_{\rm G}} \right],
\label{eq:gumbform}
\end{equation}
with
\begin{equation}
y=\frac{\nu-a}{b},
\label{eq:gumbform2}
\end{equation}
where $a$ and $b$ are location and scale parameters \citep[see, e.g.,][]{seminaire}.
The shape parameter $\gamma_{\rm G}$ characterizes the distribution:
a distribution with $\gamma_{\rm G}=0$ is 
known as having the ``Gumbel type'', 
\begin{equation}
G_{0}=\exp[-\exp(-y)],
\label{eq:asigo}
\end{equation}
 while $\gamma_{\rm G} <
0$ and $\gamma_{\rm G} >
0$ correspond respectively to forms of the ``negative Weibull type'' and the ``Fr\'echet
type''. 

Distributions given by equation (\ref{eq:gumbform}) 
have seen application to time-series data in many fields such as
climate \citep[see, e.g.,][]{Katz}, hydrology
\citep[see, e.g.,][]{hydrology},  seismology \citep[see, e.g.,][]{Cornell}, 
insurance and finance \citep[see, e.g.,][]{Embrechts}, 
etc, in predicting the incidence of extreme events from knowledge
of past data.  Here we consider applications to
two and three dimensional random fields relevant to cosmology,
but our approach is sufficiently general that extension to other
fields should not prove difficult.

In three dimensions, one is naturally
interested in the occurence of most massive clusters
in galaxy surveys \citep{BB,Holz,Cayon,Davis}, of large scale mass concentrations
\citep{Yamila} such as the Sloan Great Wall \citep{Gott}, 
or of the largest voids observed in the spatial distribution of galaxies.
In two dimensions, the most obvious application concerns the
temperature fluctuations in the Cosmic Microwave Background 
\citep[CMB,][]{Mikelsons}, in particular the analysis of the hottest hot spots \citep{Coles} and the coldest
cold spots. There are several works that suggest the existence of anomalies in current
CMB experiments \citep[see, e.g.,][]{Larson,Ayaita}, the most proeminent one being the cold spot discovered in the
temperature maps measured by the Wilkinson Microwave Anisotropy Probe
\citep[WMAP,][]{Vielva,Cruz}. 

In this work, we consider a random field in two or three dimensions, 
smoothed on some scale $\ell$, and take large
``patches'' of size $L\gg \ell$. 
The values of the field at all points inside a given patch
constitutes one sample, and the extreme value of the
field in the patch our statistic of interest (henceforth we restrict ourselves
in particular to the maxima, though the case of patch minima is exactly analogous).

Although some of the results present in this paper also apply to the
non-Gaussian case, we restrict our practical calculations to Gaussian
fields of known power spectrum and ask how to derive
analytically the distribution of patch maxima and how to
explicitly relate the results to eqs.~(\ref{eq:gumbform}) and 
(\ref{eq:gumbform2}). This effort is not new: the calculation of the 
extreme value distribution of Gaussian fields has been paid a lot of attention
by mathematicians, for time-series \citep[see, e.g.,][]{Leadbetter} but
also in larger number of
dimensions \citep[see, e.g.,][]{BR,Rosenblatt,Adler,Aldous,Piterbarg,Adler2},
leading to a number of important results.  In particular, in the very 
large patch limit, convergence to the Gumbel type distribution 
(\ref{eq:asigo}) was rigorously demonstrated \citep[e.g.,][]{BR}.
Thus, many of the results derived in this article can 
be found in the mathematical literature,  but we recall them for sake of
completeness as they are needed to understand how 
the statistics behave in various regimes.

The method we employ to estimate the extreme value statistics relies
on a local maxima approach and was
already utilised in a more rigorous mathematical set up
\citep[e.g.,][]{Adler2}. The central point is the observation that the
probability of the patch maximum being below some specified
density threshold is exactly the probability of encountering zero
points {\em above} that threshold.
If we also identify the highest point in the patch with the
field highest {\em peak} there (an
assumption which is in fact non-trivial in general; see section \ref{sec:fundass})
then the problem is reduced to that of finding the void probability for peaks
as a function of threshold. 

The distribution and clustering statistics of 
peaks has achieved a good deal of attention in the astrophysical
literature,  in part due to their role as the nucleation points of rich galactic clusters
when the field in question is that of matter overdensities.
Combining results obtained previously for the peak
abundances and their correlation functions allows us to predict the
void probability for both two and three dimensional fields, and hence the
extreme value statistics.  Once again, although most of the results
can be found in the mathematical literature, 
the key novelty of the present paper lies in the derivation and test of
approximate predictions in an intermediate regime where the patch size is
large enough compared to the coherence length of the field, but 
{\em not so large} that either the asymptotic limit expected for
Gaussian random fields \citep{BR}, (\ref{eq:asigo}), or the Poisson
regime \citep{Aldous} have been reached.

This paper is organized as follows. Section \ref{sec:theory} sets up
the general framework, with a few definitions followed by general
results. In particular, the Gumbel statistics is related to the void
probability, which is expressed in terms of average number density
of peaks above some threshold and their $N$-point correlation
functions. Section \ref{sec:gaussianfield} focuses on the Gaussian
field case, where a general estimate taking into account
full clustering of the peaks is performed 
and explicit asymptotic formulae are derived and
related to equation (\ref{eq:gumbform}). Convergence to 
(\ref{eq:asigo})  for Gaussian random fields is recovered when
the patch size tends to infinity. An explicit link to
the Euler characteristic is also established, in agreement
with the literature. In section \ref{sec:mes}, we test the 
theoretical predictions against numerical experiments in the two-dimensional case. 
Finally, section \ref{sec:discussion}
summarizes the results obtained in this article and discusses their  
generalization to non-Gaussian fields. 
\section{Theory}
\label{sec:theory}
\subsection{Definitions}
We consider, in a $D$-dimensional space with $D=2$ or 3, a random field $\delta(x)$
of zero average. We suppose that this field is statistically
stationary (invariance of the $N$-point correlation functions by
translation) and isotropic (invariance of the $N$-point correlation
functions by rotation).

\subsubsection{Smoothing window}
This field is smoothed with a window of size $\ell$:
\begin{equation}
\delta_\ell=\delta * W_\ell(x).
\end{equation}
For instance, the Gaussian smoothing window  $W_\ell(x) \propto
\exp(-x^2/2\ell^2)$, which we shall
chose for all practical calculations, reads in Fourier space
\begin{equation}
W_\ell(k)=\exp[-(k\ell)^2/2].
\end{equation}
The top hat smoothing window will be needed as well, on scales
$L \gg \ell$. In 3D, it is a sphere of radius $L$ which reads in
Fourier space:
\begin{equation}
W_L(k)=3 [ \sin(kL)-kL \cos(k L)]/(kL)^3.
\end{equation}
In 2D, it is a disc of radius $L$ which reads in
Fourier space:
\begin{equation}
W_L(k)=\frac{2 J_1(kL)}{kL},
\end{equation}
where $J_1$ is the Bessel function of the first kind and of first
order:
\begin{equation}
J_1(x)=\frac{1}{\pi} \int_0^\pi \cos[y-x \sin(y)] {\rm d} y.
\end{equation}
\subsubsection{Gumbel statistics}
From now on we measure the height of the field using the
the density contrast in units of its standard deviation,
\begin{equation}
\nu \equiv \frac{\delta_{\rm \ell}}{\sigma_0},
\end{equation}
with
\begin{equation}
\sigma_0^2 \equiv
\langle \delta_{\ell}^2 \rangle.
\end{equation}
We consider a large spherical patch of size $L$ at random position $x_0$, $L \gg \ell$ and
measure in that patch the maximum value of the smoothed density field:
\begin{equation}
\nu_{\rm max} \equiv \max\left\{ \nu(x); |x-x_0| \leq L \right\}.
\label{eq:trugumb}
\end{equation}
The goal is to study the Gumbel statistics, i.e. the probability distribution function
$p_{\rm G}(\nu_{\rm max}){\rm d}\nu_{\rm max}$ of the values of $\nu_{\rm max}$ when we
choose an infinite number of random realizations of $x_0$. This distribution contains a dependence
on the choice of smoothing and on  the size of the patch,
\begin{equation}
p_{\rm G}(\nu_{\rm max})=p_{\rm G}(\nu_{\rm max},\ell,L),
\end{equation}
which we leave implicit in the remainder of the paper.

\subsection{General results}
In a sufficiently non-degenerate field $\delta_\ell$, the set of local maxima
--the peaks of the density field-- is a discrete ensemble of points of positions $p_i$
and density $\nu_i=\delta_i/\sigma_0$. We shall assume that
this property is valid in all the subsequent calculations.

\subsubsection{Fundamental assumption}
\label{sec:fundass}
The fundamental assumption we make is that the maximum
of the density in a patch can be approximated by the 
density at the highest peak contained in the patch:
\begin{equation}
\nu_{\rm max} \simeq {\bar \nu}_{\rm max} \equiv\max\left\{ \nu_i, |p_i-x_0| \leq L
\right\}.
\label{eq:appgumb}
\end{equation}
This assumption is valid only in the regime where $L \gg \ell$.
Indeed, there can be local maxima outside the patch but sufficiently
close to its edge
such that the density measured at a point on the edge of the patch
is larger than the maximum density measured in the set of peaks
contained in the patch. In other words, if we consider the population
of local maxima of the density field defined inside the $D-1$ dimensional manifold
given by the border of the patch,  of densities ${\hat \nu}_j$, then
we have, in reality,\footnote{See \cite{Adler2} for a rigourous
  formulation corresponding to a more general patch shape than just a sphere.}
\begin{equation}
\nu_{\rm max} =\max( {\bar \nu}_{\rm max}, {\hat \nu}_j)
\geq {\bar \nu}_{\rm max}.
\end{equation}
Obviously, one expects $\nu_{\rm max}$ to approach 
${\bar   \nu}_{\rm max}$ as the ratio $L/\ell$ increases and the ratio
of the patch volume to area near its edge decreases. 
\subsubsection{General expression of the Gumbel statistics}
\label{sec:gene}
Let us define the cumulative Gumbel distribution by
\begin{equation}
P_{\rm G}(\nu)\equiv {\rm Prob.}(\nu_{\rm max} \le \nu)
\equiv \int_{-\infty}^{\nu}p_{\rm G}(\nu_{\rm max}) {\rm d}\nu_{\rm max}.
\end{equation}
Such a probability, given the assumptions of \S~\ref{sec:fundass}, is also
the probability that none of the local maxima contained in the patch
are above the threshold. In other words, if we consider the population
of local maxima satisfying $\nu_i > \nu$, none of them
belongs to the patch. This happens with a probability $P_0(\nu)$, 
where $P_0$ is the probability of finding no maxima with normalized
density larger than  $\nu$ inside a spherical cell or a disc of radius $L$:
\begin{equation}
P_{\rm G}(\nu)=P_0(\nu),
\label{eq:gumbp0}
\end{equation}
hence
\begin{equation}
p_{\rm G}(\nu)=\frac{{\rm d}P_0}{{\rm d}\nu}.
\label{eq:dgumbdp0}
\end{equation}
The calculation of such a void probability can be performed using
standard count-in-cell formalism if the number density 
$n(\nu_i > \nu)$ and the connected $N$-point
correlations functions,\footnote{The connected $N$-point correlation
functions are equal to zero for a Gaussian field if $N \geq 3$.}
$\xi_N^{\rm p}(x_1,\cdots,x_N)$, of local maxima
above the threshold, are known \citep{White,Fry,BS,SzapudiSzalay}. 

In particular, one can define the averaged correlations over a cell
of size $L$ and volume $V=(4\pi/3) L^3$ or area $V=\pi L^2$,
\begin{equation}
{\bar \xi}_N^{\rm p}(L) \equiv \frac{1}{V^N} \int_{V} {\rm d}^Dx_1 \cdots {\rm d}^D x_N
\xi_N^{\rm p}(x_1,\cdots,x_N),
\end{equation}
the normalized cumulants,
\begin{equation}
S_N^{\rm p} (L)\equiv \frac{ {\bar \xi}_N^{\rm p}(L)}{{\bar
    \xi}_2^{\rm p}(L)^{N-1}}, \quad S_1^{\rm p}\equiv
S_2^{\rm p}\equiv 1,
\end{equation}
and 
\begin{equation}
N_{\rm c} \equiv n V {\bar \xi}_2^{\rm p}(L).
\end{equation}
Each of these expressions contains an implicit $\nu$-dependence.
The number $N_{\rm c}$ represents the typical number of peaks above the
threshold per overdense patch in excess to the 
average. It measures the deviation from a pure Poisson distribution
due to clustering. 

With these definitions,   the void probability can be written
\begin{equation}
P_0(\nu_{\rm  max})=\exp\left[-n V \sigma(N_{\rm c}) \right],
\label{eq:fullp0}
\end{equation}
with
\begin{equation}
\sigma(y) = \sum_{N \geq 1} (-1)^{N-1} \frac{S_N^{\rm p}}{N!} y^{N-1}.
\label{eq:sigmadef}
\end{equation}
The challenge is now to relate the statistical properties of the
local maxima to that of the underlying density field. This is
made difficult by the fact that the void probability depends
on the full hierarchy of correlations up to any order: in
particular one has to relate the $N$-point correlation functions of
the peaks to the $N$-point correlation functions of the underlying
density field. We denote the latter by $\xi_N(x_1,\cdots,x_N)$, and
similarly the averaged $N$-point correlation functions 
of the density field by ${\bar \xi}_N(L)$.

This exercise
has been performed in detail for random Gaussian fields 
by \cite{BBKS} (hereafter BBKS) and by
\cite{BE} (hereafter BE)
 in the 3D and the 2D cases respectively, extending 
earlier calculations of \cite{Kaiser} and \cite{PW}.
Note that these latter did not consider statistics of peaks, but more
generally 
of regions of $\delta_\ell$ above the density threshold. However, in the rare
event regime considered here, the two approaches should 
become equivalent (this is discussed in
detail in BBKS). 

The non Gaussian case has been examined as well for
a quite general class of hierarchical models by \cite{BernardeauSchaeffer} 
(hereafter BS).  The statistics under consideration  in that work 
was that of overdense cells
of size $\ell$ and not of peaks of the density field smoothed
with a Gaussian window of size $\ell$.  Again, the approach of
BS should give the same results as those
obtained for peaks in the rare event regime. 

\subsubsection{Asymptotic expression of $\sigma(y)$}
\label{sec:sigma}
A fundamental result of the calculations of BS is that in the high
peak limit, i.e. $\nu \gg 1$, 
and at large enough separations, i.e.
\begin{equation}
\frac{\xi_2(x_i,x_j)}{\sigma_0^2} \ll 1,
\label{eq:xico}
\end{equation}
then, 
\begin{equation}
\xi_N^{\rm p}(x_1,\cdots,x_N) \simeq \sum_{\rm trees} \sum_{\rm labels} \prod_{\rm links} \xi_2^{\rm p}(x_i,x_j),
\label{eq:closure}
\end{equation}
in the notation of these authors. This expression is valid at least in the
framework of the minimal hierarchical-tree model. 
The trees refer to ensembles of distinct pair associations
of  elements in the ensemble $\{1,\cdots,N\}$ such that a fully connected
structure containing exactly the $N$ elements is constructed without any loop. The labels
take into account all the possible combination of elements in $\{1,\cdots,N\}$ that lead
to the same tree topology. In each tree topology, there are always $N-1$ links,  by definition.
The total number of combinations of all the trees and the labels yields $N^{N-2}$ possibilities.
Figure 1 of BS can be examined to understand the process. For instance, for the 4-point correlation
function there are 2 tree topologies: (i) the 'star' where one point
is connected to all the others, and (ii) the 'line' where one point is
connected to one or two other depending on its position (at the end or
in the middle).
There are 4 possible labelings of the star and 12 possible labelings for the line. 

Equation (\ref{eq:closure}) also applies to Gaussian fields, independently of
the shape of the smoothing window, at least if $\nu \gg 1$ and the following
condition, more restrictive than (\ref{eq:xico}) holds:
\begin{equation}
\nu^2 \frac{\xi_2(x_i,x_j)}{\sigma_0^2} \ll 1.
\label{eq:validd}
\end{equation}
Indeed, the unconnected part of the $N$-point correlation function (the moment), $\mu_n$, is given by
\begin{equation}
\mu_n=\prod_{i>j}\ [\xi_2^{\rm p}(x_i,x_j)+1]
\label{eq:munexp}
\end{equation}
in the high threshold regime \citep{PW}.
Extracting the connected part from this expression consists exactly in extracting the ensemble
of distinct pairs associations in $\{1,\cdots,N\}$ such that the corresponding topology is fully connected. In the large
separation limit, i.e. at leading order in $\xi_2^{\rm p}(x_i,x_j)$,
or equivalently if the condition (\ref{eq:validd})
is verified, only the tree topologies remain (because
they correspond to the minimum power in $\xi_2^{\rm p}$ while being fully connected), and
each label for each tree is given the same weight in equation (\ref{eq:munexp}), hence leading to equation (\ref{eq:closure})
in that regime. 

Equation (\ref{eq:closure}), reads, after volume averaging in a sphere
of radius $L$ (BS)
\begin{equation}
S_N^{\rm p}(L) \simeq N^{N-2}.
\label{eq:cumuhie}
\end{equation}
This result applies as well to the general tree-hierarchical model (BS). 
This means that the function $\sigma$ defined
in equation (\ref{eq:sigmadef}) reads (BS)
\begin{equation}
\sigma(y)=\left(1+\frac{1}{2} \theta \right) e^{-\theta}, \quad \theta
e^\theta=y.
\label{eq:sigas}
\end{equation}
Note that when $N_{\rm c} \ll 1$, which occurs at some point for a
large enough value of $\nu$ for which there are very few
peaks above the threshold in average per patch,
\begin{equation}
\sigma(N_{\rm c}) \simeq 1- N_{\rm c}/2
\end{equation}
by definition (eq.~\ref{eq:sigmadef}). 
Therefore, even though we expect the low end tail of the Gumbel
statistics to be affected by the potential crudeness of our approximation of the function
$\sigma(y)$, the high end tail should still be quite well described. 
\section{The Gaussian field case}
\label{sec:gaussianfield}
Once the function $\sigma(y)$ is specified, one needs to carry out the
calculation of the number density of peaks above the threshold as well
as their averaged two-point correlation function. The detailed
expressions can be found for a Gaussian field in BBKS and in BE.

\subsection{Shape parameters of the power spectrum: $\gamma$ and $R_\star$}
The important parameters that control the number density of peaks
above threshold $\nu$ and their two point
correlation function are the moments
\begin{equation}
\sigma^2_j \equiv \int \frac{k^{D-1} {\rm d}k}{2\pi^{D-1}} P(k)W_\ell^2(k)
k^{2j},
\label{eq:sigj3d}
\end{equation}
where $D=2$ or 3 is the dimension of the space considered.
Then BBKS and BE define the coherence parameter $\gamma$
\begin{equation}
\gamma\equiv \frac{\sigma_1^2}{\sigma_0 \sigma_2}
\end{equation}
and the scale length $R_\star$ by
\begin{equation}
R_\star=\sqrt{D} \frac{\sigma_1}{\sigma_2}.
\end{equation}

For a Gaussian smoothing window and a scale free power spectrum $P(k)$ given by
\begin{equation}
P(k)=A k^n,
\end{equation}
the integrals in equation (\ref{eq:sigj3d}) 
can be performed analytically, yielding the simple expressions
\begin{equation}
\sigma_0^2=\left( \frac{\ell}{\ell_0}\right)^{-(n+D)}, \ \ 
\ell_0=\left[ \frac{A}{4\pi^{D-1}} \Gamma\left(\frac{n+D}{2}
  \right)\right]^{1/(n+D)},
\label{eq:sigmazero}
\end{equation}
\begin{equation}
\gamma=\sqrt{\frac{n+D}{n+D+2}}, \quad R_\star=\sqrt{\frac{2D}{n+D+2}} \ell,
\label{eq:gamma_rstar}
\end{equation}
valid for $n>-D$.\footnote{The case $D$=3 was derived in BBKS, with the expression
for the correlation length $\ell_0$ given in e.g. \cite{Lokas}.}

For a top-hat smoothing the scaling law (\ref{eq:sigmazero}) remains
valid with a different correlation length, which for the 3D case
can be written \citep[see, e.g.,][]{Lokas}
\begin{equation}
\ell_0=\left\{ \frac{9A\ \Gamma[(n+3)/2]\ \Gamma[(1-n)/2]}{8
    \pi^{3/2}\ \Gamma[1-n/2]\ \Gamma[(5-n)/2]}\right\}^{1/(n+3)}.
\end{equation}
On the other hand, we did not find any simple analytic expression of the
correlation length for a top-hat smoothing in 2D.

\subsection{Number density of peaks}
The number density of peaks above the threshold is
\begin{equation}
n(\nu)=\int_{\nu}^{\infty} d\nu' {\cal N}(\nu').
\end{equation}
This integral is easily performed numerically, using
the fact that the function ${\cal N}(\nu)$ is given by
\begin{equation}
{\cal N}(\nu)d\nu=\frac{1}{(2\pi)^{(D+1)/2} R_\star^D} e^{-\nu^2/2}
G_D(\gamma,\gamma\nu)
\label{eq:nbbks}
\end{equation}
in $D$ dimensions,
with $G_3$ approximated by equation (4.4) of BBKS and $G_2$ given by
equation (A1.9) of BE. For completeness, we rewrite these equations
here:
\begin{equation}
G_3(\gamma,w)  \simeq \nonumber \frac{w^3-3\gamma^2 w +
  (B w^2+C_1)\exp(-A w^2)}{1+C_2\exp(-C_3w)},
\end{equation}
with
\begin{eqnarray}
A &= & \frac{5}{2(9-5\gamma^2)},\\
B & = & \frac{432}{\sqrt{10\pi} (9-5\gamma^2)^{5/2}},\\
C_1 & = & 1.84+1.13 (1-\gamma^2)^{5.72},\\
C_2 & = & 8.91 + 1.27 \exp(6.51 \gamma^2),\\
C_3 & = & 2.58 \exp (1.05 \gamma^2),
\end{eqnarray}
and
\begin{eqnarray}
G_2(\gamma,w) & = & (w^2-\gamma^2) \left\{ 1-\frac{1}{2} {\rm erfc}\left[
    \frac{w}{\sqrt{2(1-\gamma^2)} } \right] \right\} \nonumber \\
 &+ & \frac{w(1-\gamma^2)}{\sqrt{2\pi(1-\gamma^2)}} \exp\left[
   -\frac{w^2}{2(1-\gamma^2)}\right]\nonumber \\
 & + & \frac{1}{\sqrt{3-2\gamma^2}} \exp\left(
   -\frac{w^2}{3-2\gamma^2} \right) \times \nonumber \\
 &  & \left\{ 1 -\frac{1}{2} {\rm erfc}\left[ \frac{w}{\sqrt{2(1-\gamma^2)(3-2\gamma^2)}}
   \right] \right\}.
\label{eq:myG2}
\end{eqnarray}
\subsection{Correlation function of peaks}
In the large separation regime the two point correlation function of
the peaks (\ref{eq:validd}) reads, if one neglects contributions from higher order
derivatives of $\xi_2(r)$, which is a fair approximation according to 
BBKS and BE if $\xi_2$ is a power-law of negative index,
\begin{equation}
\xi_2^{\rm p}=\frac{\langle{\tilde \nu}\rangle^2}{\sigma_0^2} \xi_2,
\end{equation}
where
\begin{equation}
\langle{\tilde \nu} \rangle= \frac{\int_{\nu}^{\infty}  {\tilde \nu}(\nu')
  {\cal N}(\nu') {\rm d}\nu'}{\int_{\nu}^{\infty} {\cal N}(\nu')
  d\nu'},
\label{eq:avenu}
\end{equation}
and the effective threshold ${\tilde \nu}(\nu)$ writes
\begin{equation}
\tilde \nu= \nu - \frac{\gamma \theta}{1-\gamma^2},
\label{eq:tildenu}
\end{equation}
with $\theta$ approximated by equation (6.14) of BBKS in 3D:
\begin{equation}
\theta\simeq \frac{3(1-\gamma^2)+(1.216-0.9 \gamma^4) \exp[ -\gamma/2(\gamma
  \nu/2)^2 ]}{\sqrt{ 3(1-\gamma^2)+0.45+(\gamma\nu/2)^2}+\gamma \nu/2},
\end{equation}
 and given in 2D by
\begin{equation}
\theta=(1-\gamma^2) \frac{H(\gamma,\gamma \nu)}{G_2(\gamma,\gamma \nu)},
\label{eq:theta2D}
\end{equation}
where $H(\gamma,w) \equiv \partial G_2/\partial w$ is given by equation
(A4.7a) of BE.  For completeness,
\begin{eqnarray}
H(\gamma,w) & = & 2 w \left\{ 1 -\frac{1}{2} {\rm erfc} \left[ \frac{w}{\sqrt{2(1-\gamma^2)}}\right]
\right\} \nonumber \\
& + & \frac{4 (1-\gamma^2)^2}{(3-2\gamma^2)\sqrt{2\pi(1-\gamma^2)}}
\exp\left[ -\frac{w^2}{2(1-\gamma^2)}\right] \nonumber \\
& - & \frac{2 w}{(3-2\gamma^2)^{3/2}} \exp\left(-
  \frac{w^2}{3-2\gamma^2} \right) \times \nonumber \\
& &  \left\{ 1 -\frac{1}{2} {\rm erfc} \left[
    \frac{w}{\sqrt{2(1-\gamma^2)(3-2\gamma^2)}}\right] \right\}.
\label{eq:funcH}
\end{eqnarray}
In the large threshold limit, we simply have
\begin{equation}
\langle {\tilde \nu} \rangle \rightarrow \nu, \quad \nu \rightarrow \infty,
\end{equation}
as derived by \cite{Kaiser}. 
After averaging over volume $V$, the expression of ${\bar \xi}^{\rm p}_2(L)$ is thus simply given by
\begin{equation}
{\bar \xi}_2^{\rm p}(L)= \frac{\langle {\tilde \nu}
  \rangle^2}{\sigma_0^2} {\bar \xi}_2(L),
\label{eq:xipeakofL}
\end{equation}
where ${\bar \xi}_2(L)$ is the averaged two-point correlation function
of the underlying density field. It can be derived easily from the power spectrum
of the underlying (smoothed) density field, $\delta_{\ell}$, using
equation (\ref{eq:sigj3d}) with the top
hat window and replacing $\ell$ with $L$. The largeness of the patch
size, $L$, compared to the smoothing scale, $\ell$, should guarantee that the large
separation approximation (\ref{eq:validd}) is verified in
practice. This can be checked a posteriori by examining the
range of values of $\nu$ where $p_{\rm G}(\nu)$ is significant. 

Note that \cite{Sheth} and \cite{Desjacques} (the latter in the large separation limit) 
performed the exact calculation of the two-point correlation
function of peaks in 2D and 3D, respectively, by
taking into account corrections depending on second and fourth derivative
of the two-point correlation function of the underlying field.  
These corrections can be significant on large scales, 
where the power-spectrum of the underlying field significantly 
deviates from a power-law.  However, they get progressively
smaller with increasing threshold ${ \nu}$, and in the rare event limit in which we are
working here, they are probably irrelevant, except perhaps for the low-$\nu$ tail
of the Gumbel statistics.  Still, this assumption should be checked 
explicitly for the Cold Dark Matter case by assessing the differences
in the distribution introduced by computing the two-point correlation function of peaks with and
without them. Such an investigation is beyond the scope
of this paper, but it should be kept in mind when comparing analytic
estimates of the Gumbel statistics to real data. Also note that in this more rigorous
context,  the simple proportionality relation (\ref{eq:xipeakofL}) does not apply anymore. 

\subsection{Asymptotic regime}
\label{sec:asymptotic}
A particularly interesting case corresponds to the regime $\nu \gg 1$
and  the Poisson limit $N_{\rm c} \ll 1$, where $\sigma(N_{\rm c})
\simeq 1$. Such a regime is expected to be reached if $L/\ell$ is
sufficiently large and has been studied previously
\citep{Aldous}. However, since they will prove to be very useful to understand the
intermediate (as opposed to asymptotic) regime that we discuss later,
we recast the main results using our formalism. Here, the calculations will be facilitated by
examining the cumulative Gumbel distribution, $P_{\rm G}(\nu)$.

In the large $\nu$ regime, the number density of peaks is proportional to
the Euler Characteristic ${\cal E}$ of the underlying density field\footnote{The Euler characteristic is seen here as an alternate
  count of critical points number densities of various kinds included in the excursion
in regions with normalized density larger than $\nu$, 
see e.g. \cite{CPS,Adler2}.} 
(e.g., BBKS, BE). Just how large a value of $\nu$ is required for this to be true
depends on the level of accuracy one aims
to reach in the description of the function $P_{\rm G}(\nu)$.  For instance,
BBKS suggest $\gamma \nu > 2.5$ in 3D for a 10
percent accuracy on approximating $n(\nu)$ by the Euler
Characteristic.

With the additional assumption that $N_{\rm c} \ll
1$,  equations (4.14) of BBKS and equation (3.3) of BE read,
respectively in 3D and 2D, 
\begin{eqnarray}
P_{\rm G,3}(\nu) & \simeq & \exp( -{\cal E}_3V) \nonumber \\
 & &=\exp\left[  -U_3 (\nu^2-1)
  \exp\left(-\frac{\nu^2}{2} \right) \right],
\label{eq:asy3D}
\end{eqnarray}
\begin{eqnarray}
P_{\rm G,2}(\nu) & \simeq & \exp(-{\cal E}_2V) \nonumber \\
 & & =\exp\left[ -U_2\ \nu
  \exp\left( -\frac{\nu^2}{2} \right) \right].
\label{eq:asy2D}
\end{eqnarray}
with
\begin{equation}
U_D=\frac{\gamma^D V}{(2\pi)^{(D+1)/2} R_\star^D}.
\label{eq:UDgen}
\end{equation}
For scale free power-spectra we have
\begin{equation}
U_D=\left(\frac{4}{3} \right)^{D-2}\frac{\pi}{(2\pi)^{(D+1)/2}} \left( \frac{n+D}{2D} \right)^{D/2}
  \left( \frac{L}{\ell} \right)^D.
\end{equation}
Note that in the original derivation \citep{Aldous}, the right hand
part of equation (\ref{eq:asy3D}) contains a term scaling like $\nu^2$
and not $\nu^2-1$. Recall however that these expressions still assume $\nu$
sufficiently large compared to unity. Note also that in the very large
threshold limit, one recovers the classical result \citep[see][]{Adler,Adler2}:
\begin{equation}
1-P_{\rm G,D}(\nu \gg 1) \simeq {\cal E}_DV.
\end{equation}
In the calculations presented in \cite{Adler2}, though, the edge effects discussed in
\S~\ref{sec:fundass} are not neglected, i.e. ${\cal E}_D V$ must be in
fact viewed as the ensemble average of the Euler characteristic of the
intersection between the excursion and the patch.\footnote{Note that 
having a very accurate determination of the high $\nu$ tail of the Gumbel
distribution has been paid a lot of attention by
mathematicians and numerous methods have been employed
to do so, estimating the Euler characteristic being one amongst them
\citep[see the introduction of][for a panorama on various
methods]{Azais}.}

An interesting value of $\nu$ corresponds to 
\begin{equation}
n(\nu_\star)V=1,
\label{eq:nuimpli}
\end{equation}
 or $P_{\rm G}=1/{\rm e}$. Obviously we must have $\nu_\star$ sufficiently
large compared to unity for equations (\ref{eq:asy3D}) and (\ref{eq:asy2D}) to
hold, as well as 
\begin{equation}
N_{\rm c}(\nu_\star)=\frac{\nu_\star^2}{\sigma_0^2} {\bar
  \xi}_2(L) \ll 1,
\end{equation}
to remain in the Poisson limit.  The last condition imposes a constraint
on the size $L$ of the patch, which must be large enough 
compared to the smoothing scale $\ell$. This
obviously depends on spectral index: the ratio $L/\ell$ should 
be larger if $n$ is small since there is more power on large
scale. 

Asymptotically, $\nu_\star$ reads 
\begin{equation}
\nu_\star \simeq \sqrt{2 \ln U_D} \left[ 1+\frac{(D-1) \ln(2\ln U_D)}{
    4\ln U_D} \right].
\label{eq:nuD}
\end{equation}
This equation shows that $\nu_\star$ grows rather slowly with $L/\ell$. 

Now we compare the expressions (\ref{eq:asy3D}) and (\ref{eq:asy2D}) to
the standard cumulative form  (\ref{eq:gumbform}). 
To determine the parameters $a$, $b$ and $\gamma_{\rm G}$ of
eqs.~(\ref{eq:gumbform}) and (\ref{eq:gumbform2}), 
we perform a second order Taylor expansion near $y=0$, where 
$G_{\gamma_{\rm G}}(y=0)=1/{\rm e}$. 

At second order in $\gamma_{\rm G} y$ we have
\begin{equation}
\ln(-\ln G_{\rm \gamma_{\rm G}}) \simeq -y+\gamma_{\rm G} \frac{y^2}{2}.
\end{equation}
Similarly we have
\begin{eqnarray}
\ln(-\ln P_{G,3}) & \simeq & -\frac{a^2}{2} + \ln[ U(a^2-1)] \nonumber \\  
  & & - \frac{ab(a^2-3)}{a^2-1} y -\frac{b^2(a^4+3)}{(a^2-1)^2} \frac{y^2}{2},
\end{eqnarray}
\begin{eqnarray}
\ln(-\ln P_{G,2}) & \simeq & -\frac{a^2}{2} + \ln( U a) \nonumber \\
 & & -\frac{b(a^2-1)}{a} y - \frac{b^2(a^2+1)}{a^2} \frac{y^2}{2}.
\end{eqnarray}
Our particular choice of the expansion is convenient since it implies 
\begin{equation}
a=\nu_\star.
\label{eq:aD}
\end{equation}
Then
\begin{equation}
b_3=\frac{1}{\nu_\star} \frac{\nu_\star^2-1}{\nu_\star^2-3},
\label{eq:b3D}
\end{equation}
\begin{equation}
b_2=\frac{\nu_\star}{\nu_\star^2-1},
\label{eq:b2D}
\end{equation}
\begin{equation}
\gamma_{\rm G,3}=-\frac{\nu_\star^4+3}{\nu_\star^2(\nu_\star^2-3)^2} < 0,
\label{eq:g3D}
\end{equation}
\begin{equation}
\gamma_{\rm G,2}=-\frac{\nu_\star^2+1}{(\nu_\star^2-1)^2} < 0,
\label{eq:g2D}
\end{equation}
where the labels 2 and 3 refer to the number of dimensions
considered, $D$.
It is interesting to study the asymptotic values of these parameters
when $\ln U_D \gg 1$, hence when $\nu_\star$ is very large:
\begin{equation}
b \sim 1/\nu_\star,\quad \gamma_{{\rm G}} \sim -1/\nu_\star^2,
\end{equation}
thus
\begin{equation}
\gamma_{\rm G} y \sim \frac{\nu}{\nu_\star}-1.
\end{equation}
We can thus understand that the range of validity of the Taylor expansion
translated in terms of the $\nu$ variable is $\nu \in
[\nu_\star(1-\varepsilon),\nu_\star(1+\varepsilon)]$, where $\varepsilon$ is a
fraction of unity, for both $P_{\rm G}(\nu)$ and
$G_{\gamma_{\rm G}}(\nu)$. While $\nu_\star$ does not in general correspond
exactly to the position of the maximum of $p_{\rm G}(\nu)$
and
\begin{equation}
g_{\gamma_{\rm G}}(\nu) \equiv \frac{{\rm d}G_{\gamma_{\rm G}}}{{\rm
      d}\nu}, 
\end{equation}
it is rather close to it, and increasingly so as $L/\ell$ becomes larger. 
This means in practice that
the functions $p_{\rm G}(\nu)$ and $g_{\gamma_{\rm G}}(\nu)$ should
be well described close to their maximum by our second order Taylor
expansion in an interval corresponding 
to confidence levels up to $\sim$90-95\%. Hence, the functions
$p_{\rm G}(\nu)$ and $g_{\gamma_{\rm G}}(\nu)$ should match
each other quite well in that interval, but not in the tails,
especially in the large $\nu$ one.  

Notice as well that $\gamma_{\rm G}$ is negative, as measured experimentally
by e.g. \cite{Mikelsons} and that it converges to zero, so
that the function $P_{{\rm G}}(\nu)$ converges to eq.~(\ref{eq:asigo})
as demonstrated more rigorously by e.g. \cite{BR} \citep[see
also][]{Rosenblatt}.  

This asymptotic result was first exploited in cosmology by \cite{Coles} to
analyze the hottest hot spots in temperature fluctuations of the 
CMB. However, it is important to
notice that convergence to the form (\ref{eq:asigo}) is rather slow.
On the other hand, the form (\ref{eq:gumbform}) with values of $a$, $b$ and
$\gamma_{{\rm G}}$ given by equations (\ref{eq:aD}), (\ref{eq:b3D}),
(\ref{eq:b2D}), (\ref{eq:g3D}) and (\ref{eq:g2D}) along with
implicit equation (\ref{eq:nuimpli}) to determine $\nu_\star$
remains always a good fit of equations (\ref{eq:asy3D}) and
(\ref{eq:asy2D}) in the 90-95\% confidence region, but again, not in the
tails.  These analytical results will be illustrated explicitly in \S~\ref{sec:mes}.

\section{Measurements}
\label{sec:mes}
To check the validity of the theoretical calculations, we performed
numerical experiments in the 2D case. We generated scale free random Gaussian
fields on sets of 100 realizations on a grid 
of size $4096^2$ for each value of the spectral index we considered,
$n=0$, $-0.5$, $-1$ and $-1.5$. Smoothing was performed with a Gaussian
window of size $\ell=5$ pixels and 400 non-overlapping circular
patches of radius $L=100$ pixels were extracted from each realization,
amounting to a grand total of $40000$ patches to measure
$p_{\rm G}(\nu)$ for each value of $n$. 

\begin{figure*}
\centerline{\hbox{
\includegraphics[width=0.5\textwidth]{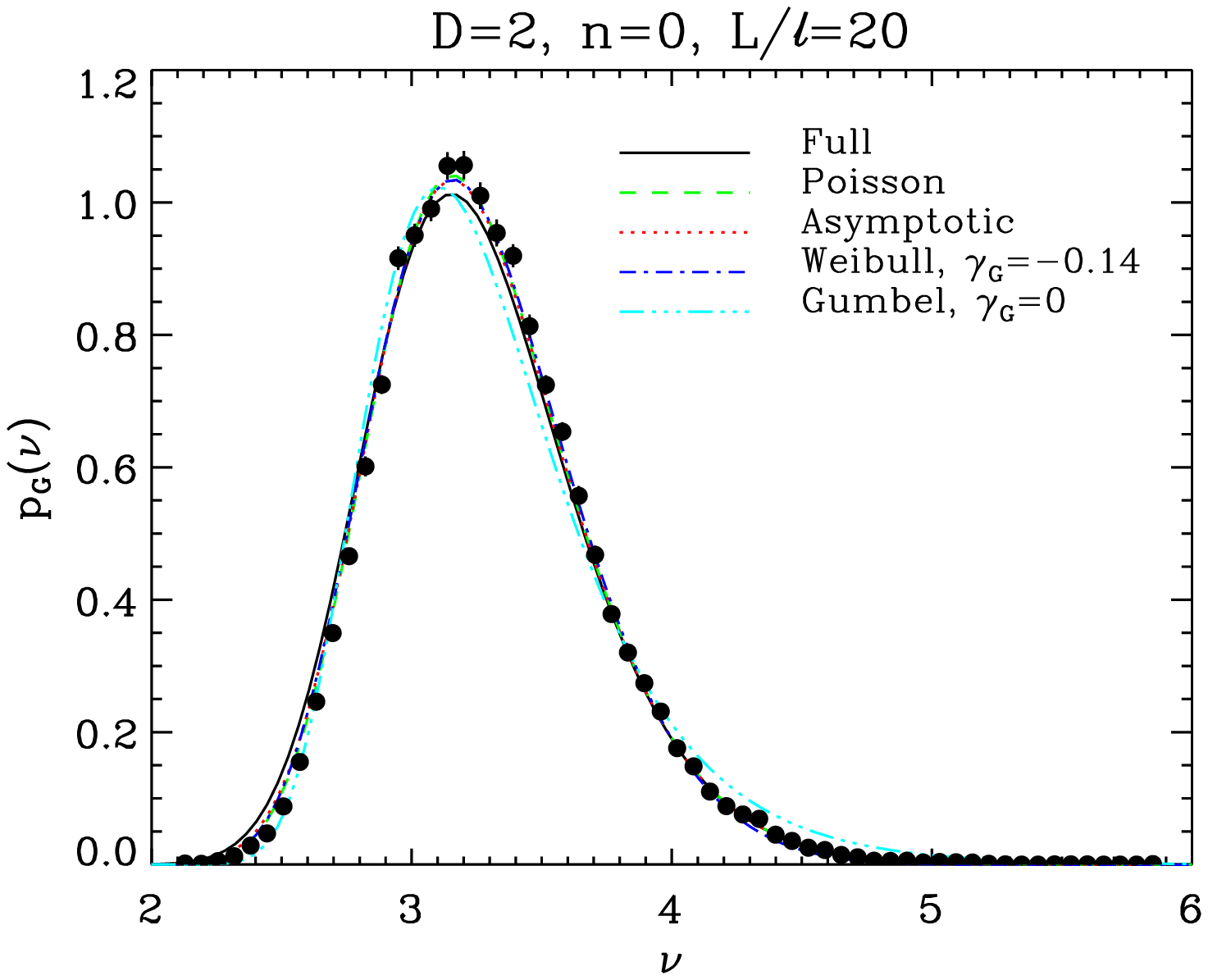}
\includegraphics[width=0.5\textwidth]{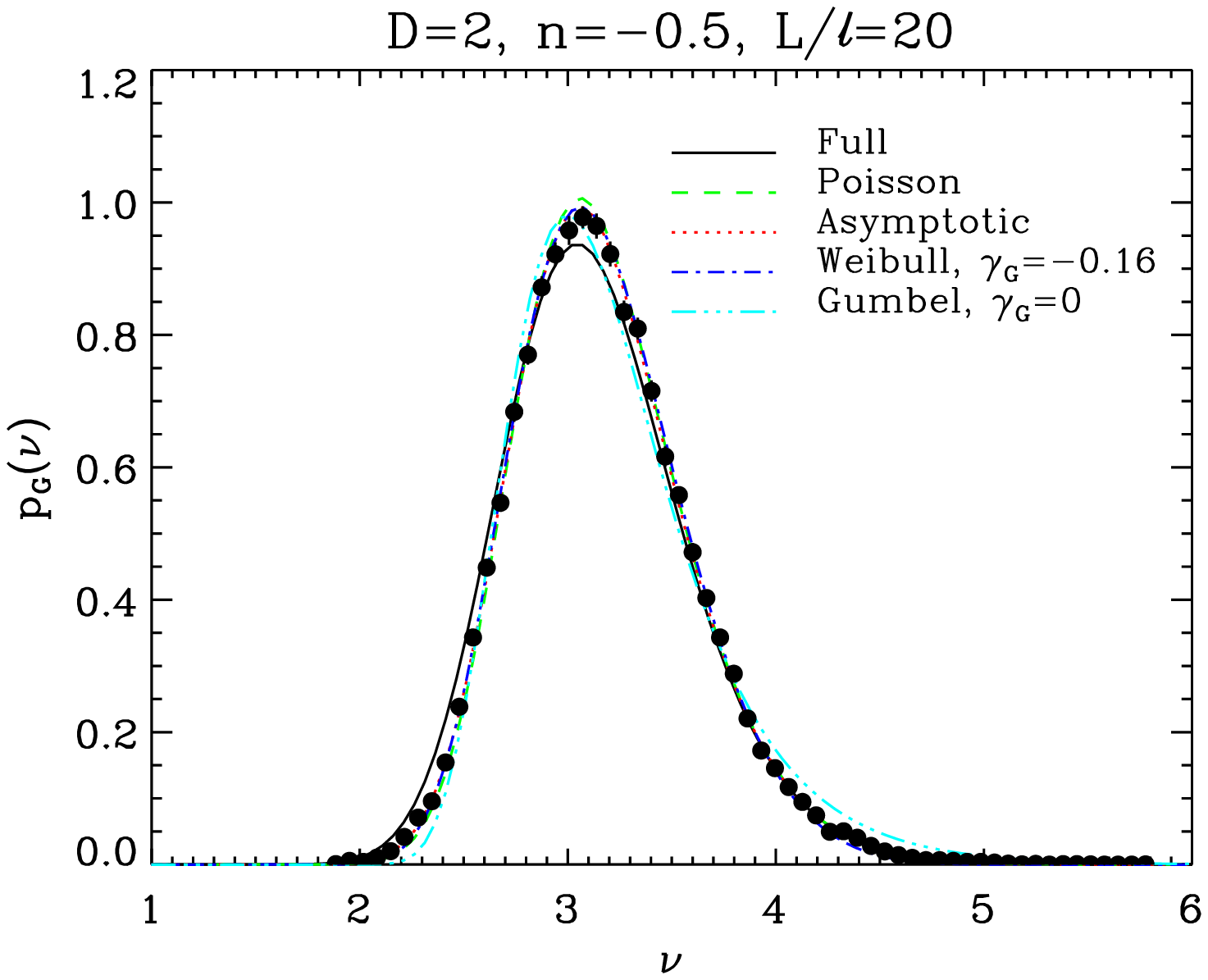}
}}
\centerline{\hbox{
\includegraphics[width=0.5\textwidth]{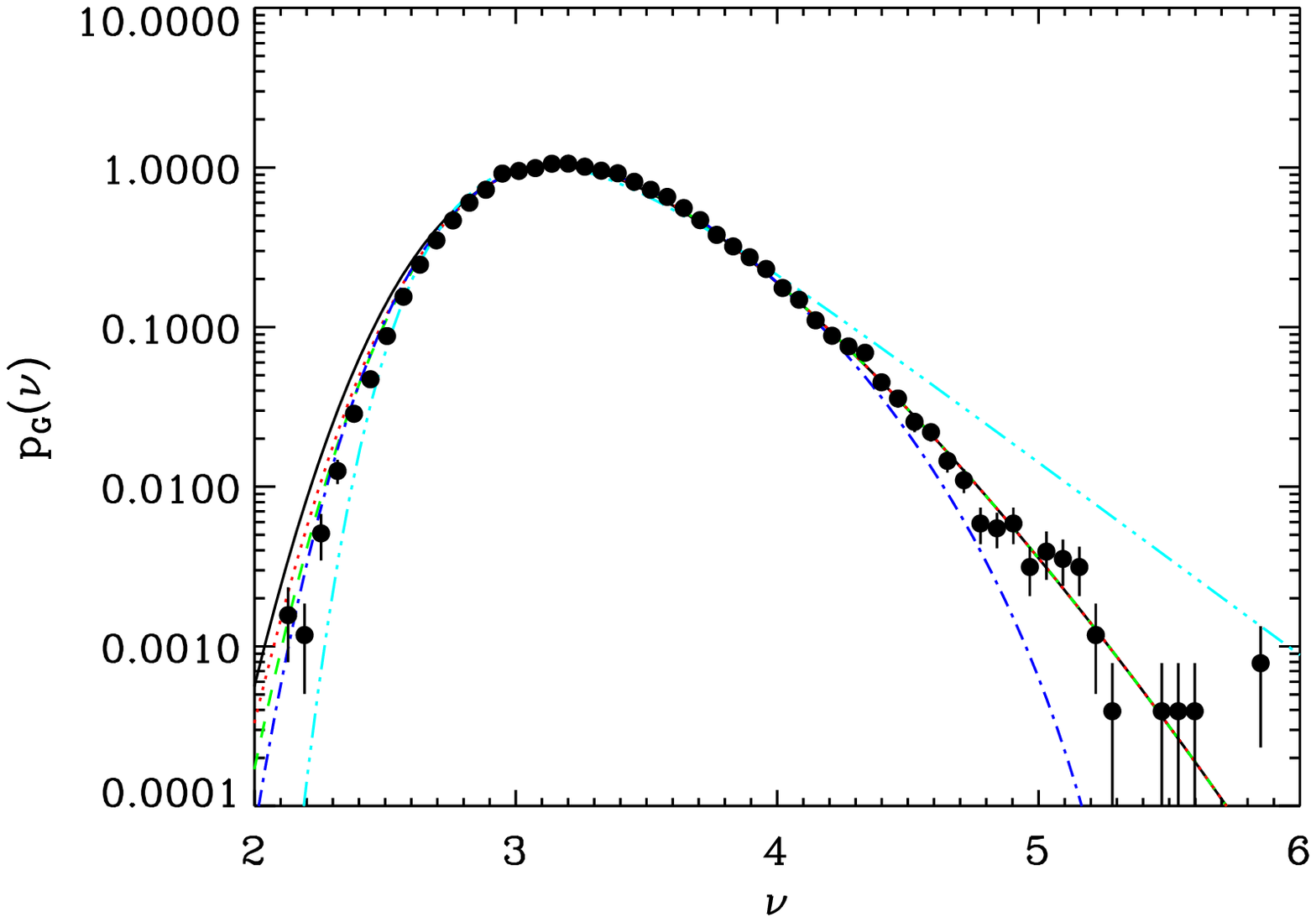}
\includegraphics[width=0.5\textwidth]{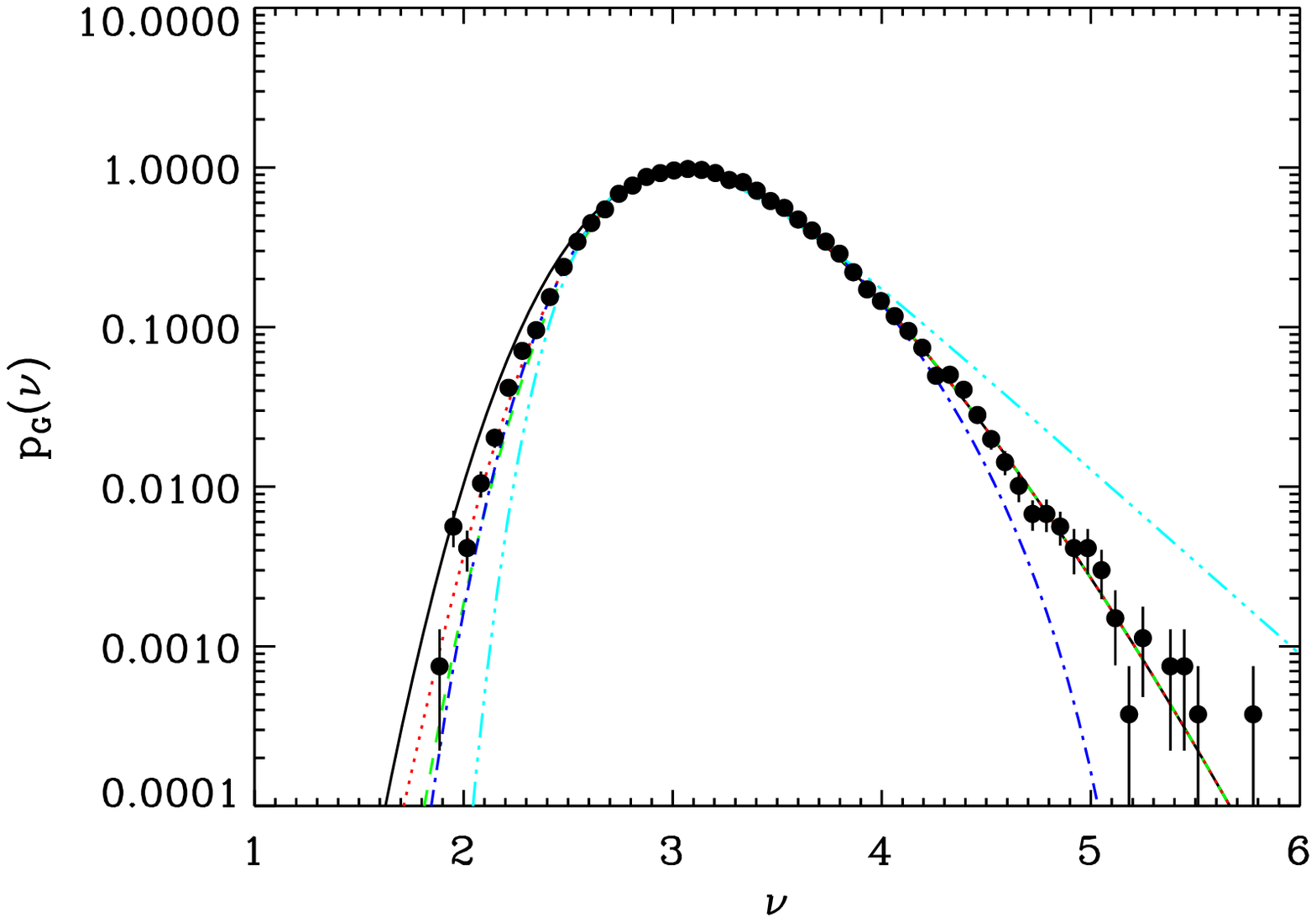}
}}
\centerline{\hbox{
\includegraphics[width=0.5\textwidth]{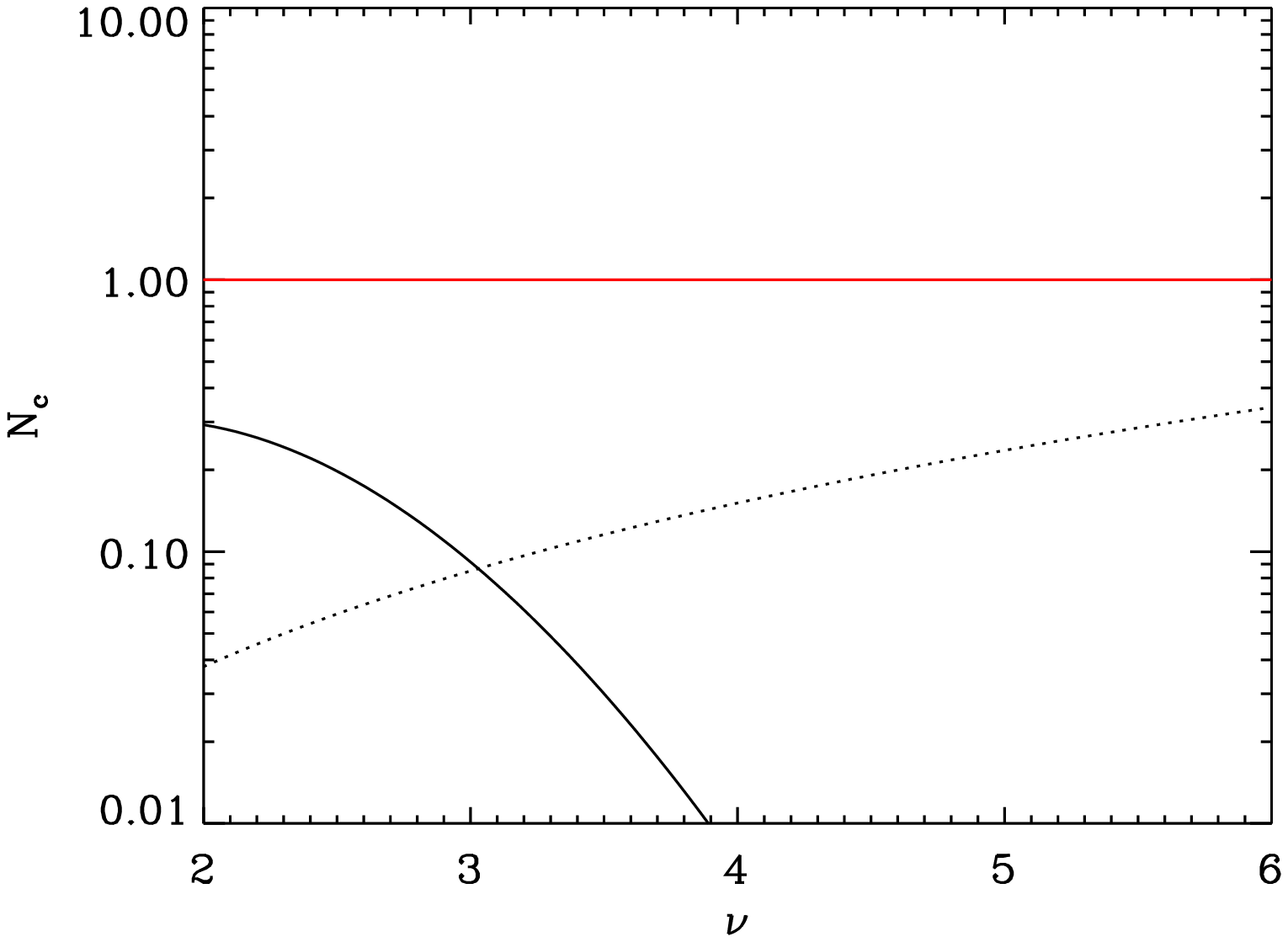}
\includegraphics[width=0.5\textwidth]{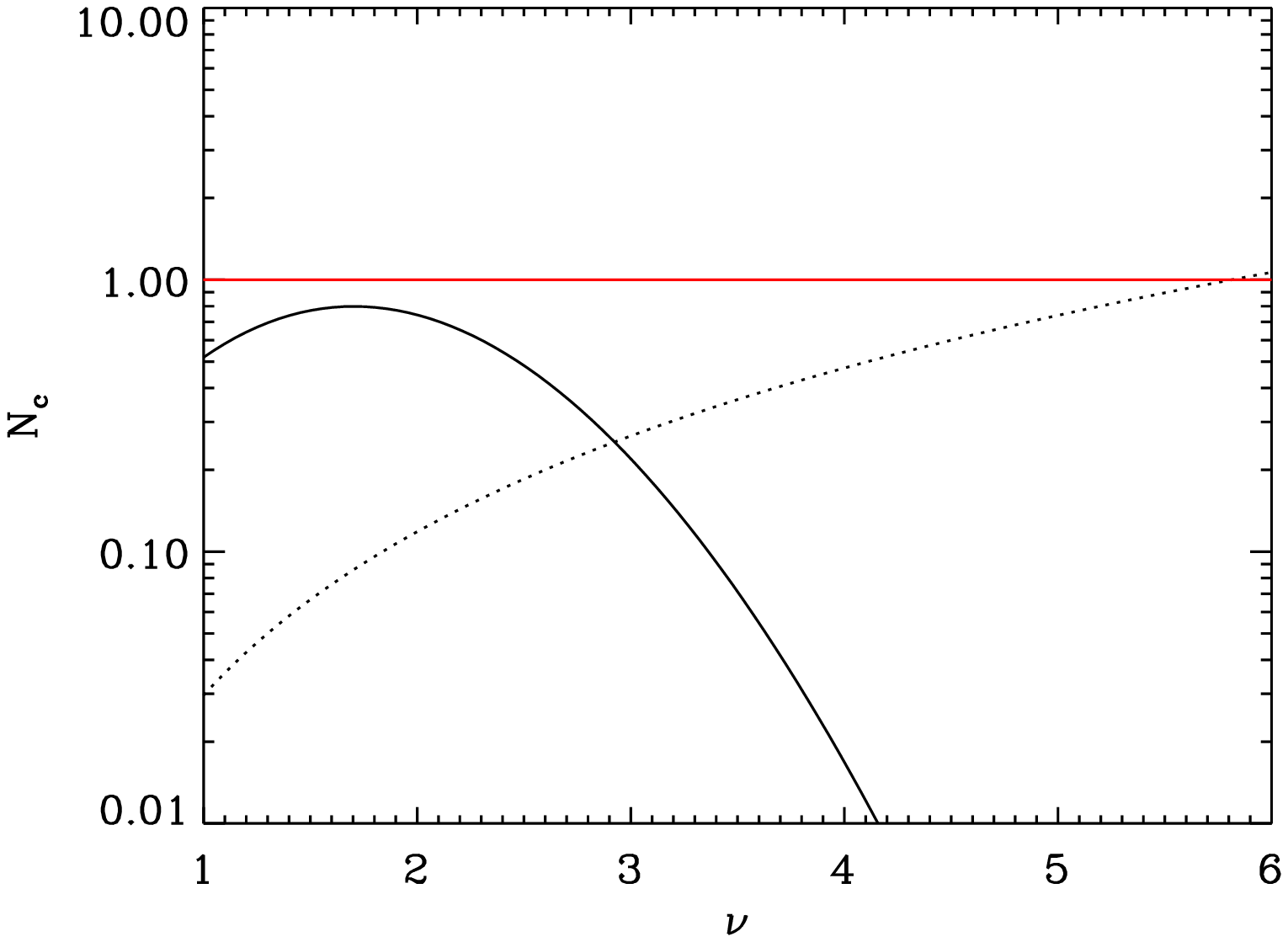}
}}
\caption[]{The Gumbel statistics measured in the case $L/\ell=20$ for
  2D random Gaussian fields with power-spectra $P(k) \propto k^n$. The
spectral index, $n=0$ and $n=-0.5$, is indicated on each panel. The values
$n=-1$ and $n=-1.5$ are examined on Fig.~\ref{fig:pdfs2}.

{\em First and second row of panels:} $p_{\rm G}(\nu)$
and its logarithm as a function of $\nu$. 
The symbols correspond to the measurements in simulated data as described in the text.
 Vertical errorbars show $1\sigma$ errors calculated from 100 independent
 realizations of the field. As indicated on the top panels, the solid
curves correspond to our theoretical prediction
(eqs.~\ref{eq:gumbp0}, \ref{eq:fullp0}, \ref{eq:sigas}, \ref{eq:nbbks}, \ref{eq:myG2},
\ref{eq:avenu}, \ref{eq:tildenu}, \ref{eq:theta2D}, \ref{eq:funcH}); the short dashed ones are
the same but assume that peaks are unclustered (Poisson limit, or
$N_{\rm c}=0$, equivalently $\sigma=1$ in eq.~\ref{eq:fullp0}, but
still eqs.~\ref{eq:nbbks} and \ref{eq:myG2} to determine the peak abundance); the dotted ones further assume 
that the number density of peaks in the
excursion is approximated by the Euler Characteristic
(eq.~\ref{eq:asy2D}); 
the dot-dashed curves give the form (\ref{eq:gumbform}) fitted
on the dotted curves, with the value of $\gamma_{\rm G}$ obtained from
matching the Taylor expansion discussed in \S~\ref{sec:asymptotic}
(eqs.~\ref{eq:nuimpli}, \ref{eq:aD}, \ref{eq:b2D}, \ref{eq:g2D});
Finally, the last curves (3 dots-one dash repeated) 
correspond to the asymptotic behavior
(\ref{eq:asigo}) expected when the ratio $L/\ell$ approaches
infinity: they are the same as the dot-dashed curves but with $\gamma_{\rm
  G}=0$.  

{\em Third row of panels:} 
$N_{\rm c}$ (solid curves) and
${\tilde \xi}_2^{\rm p}\equiv 
\nu^2 {\bar \xi}_2(L)/\sigma_0^2$ (dotted curves) as functions of $\nu$. When $N_{\rm c} \ga
1$, one expects the effect of the clustering of peaks to become
significant. On the other hand, when ${\tilde \xi}_2^{\rm p} \ga 1$,
our description of the $N$-point correlation
functions of peaks becomes inaccurate, but this only has
a significant impact on the analytical calculation of $p_{\rm G}(\nu)$
if $N_{\rm c} \ga 1$. Note that the intersection of the solid
curve and the dotted curve is expected to be in the vicinity of the maximum
of $p_{\rm G}(\nu)$. 
\label{fig:pdfs}}
\end{figure*}

\begin{figure*}
\centerline{\hbox{
\includegraphics[width=0.5\textwidth]{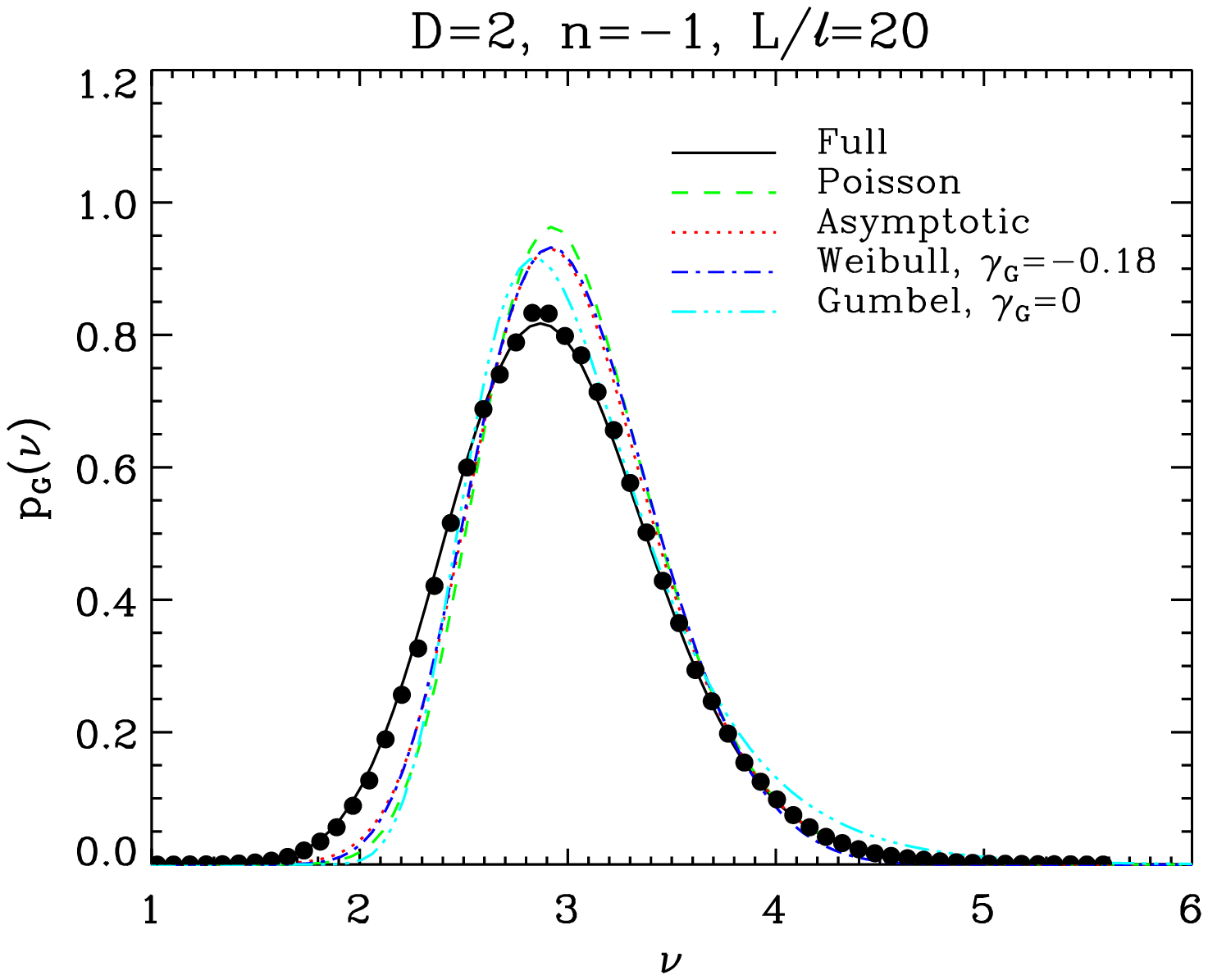}
\includegraphics[width=0.5\textwidth]{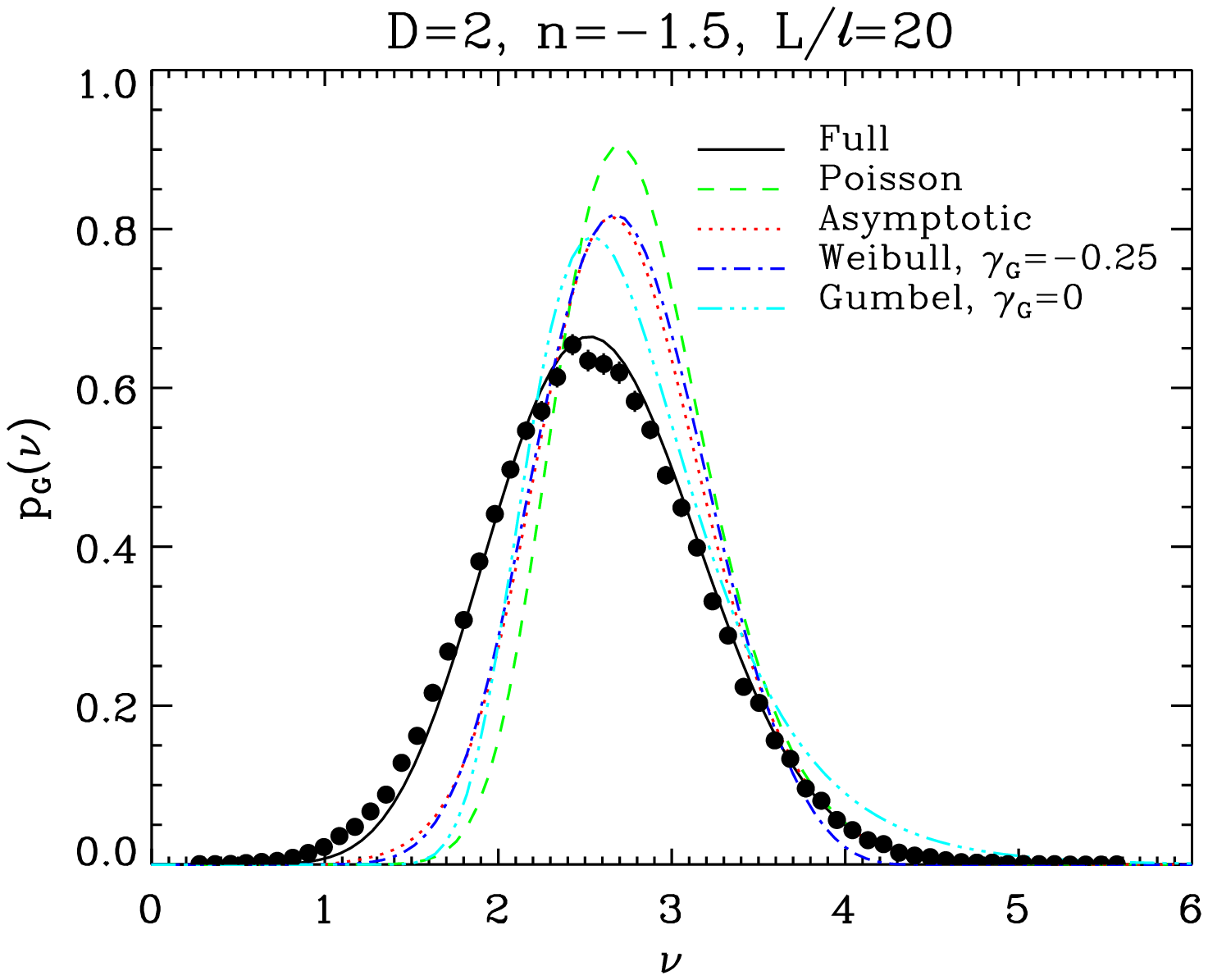}
}}
\centerline{\hbox{
\includegraphics[width=0.5\textwidth]{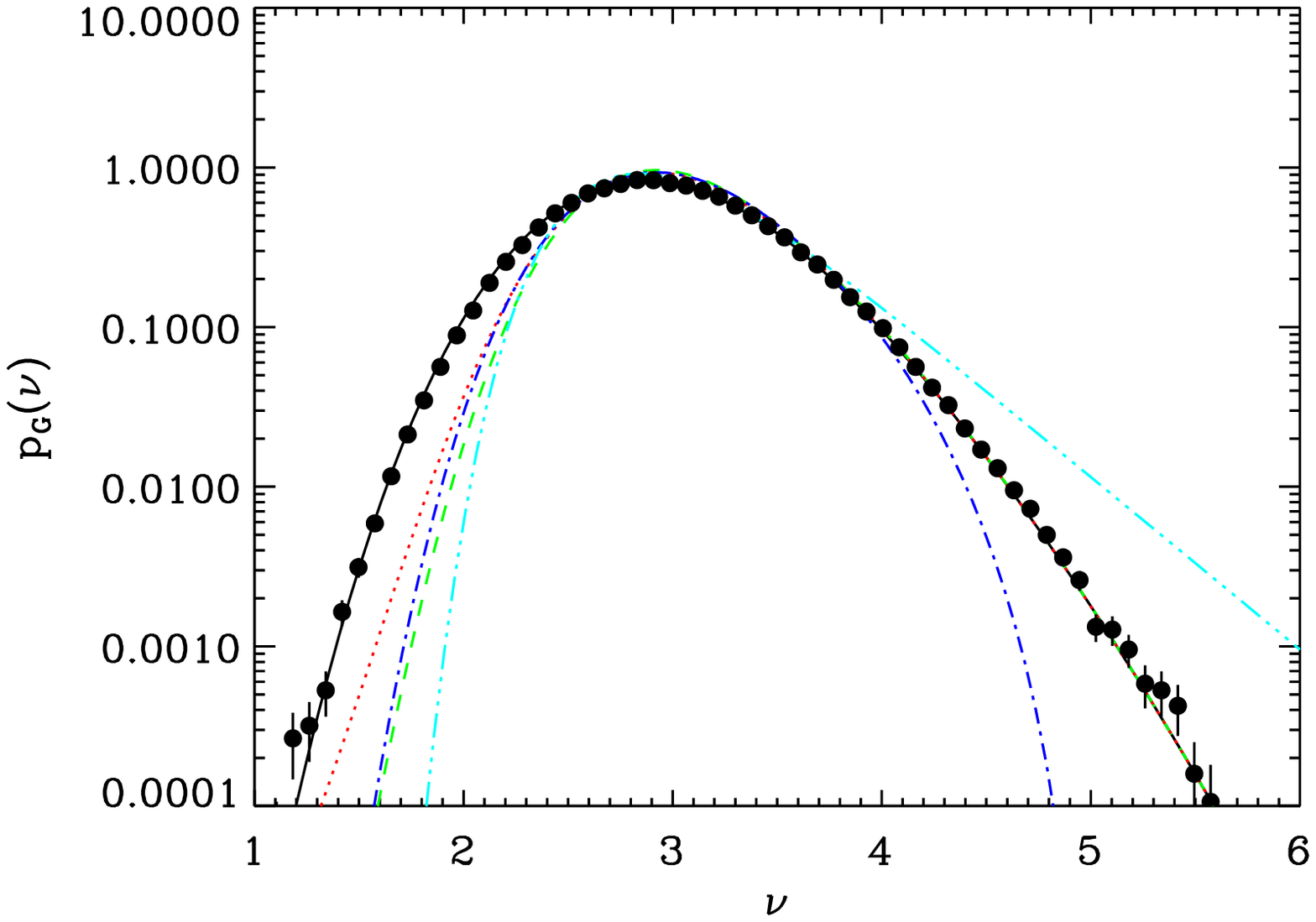}
\includegraphics[width=0.5\textwidth]{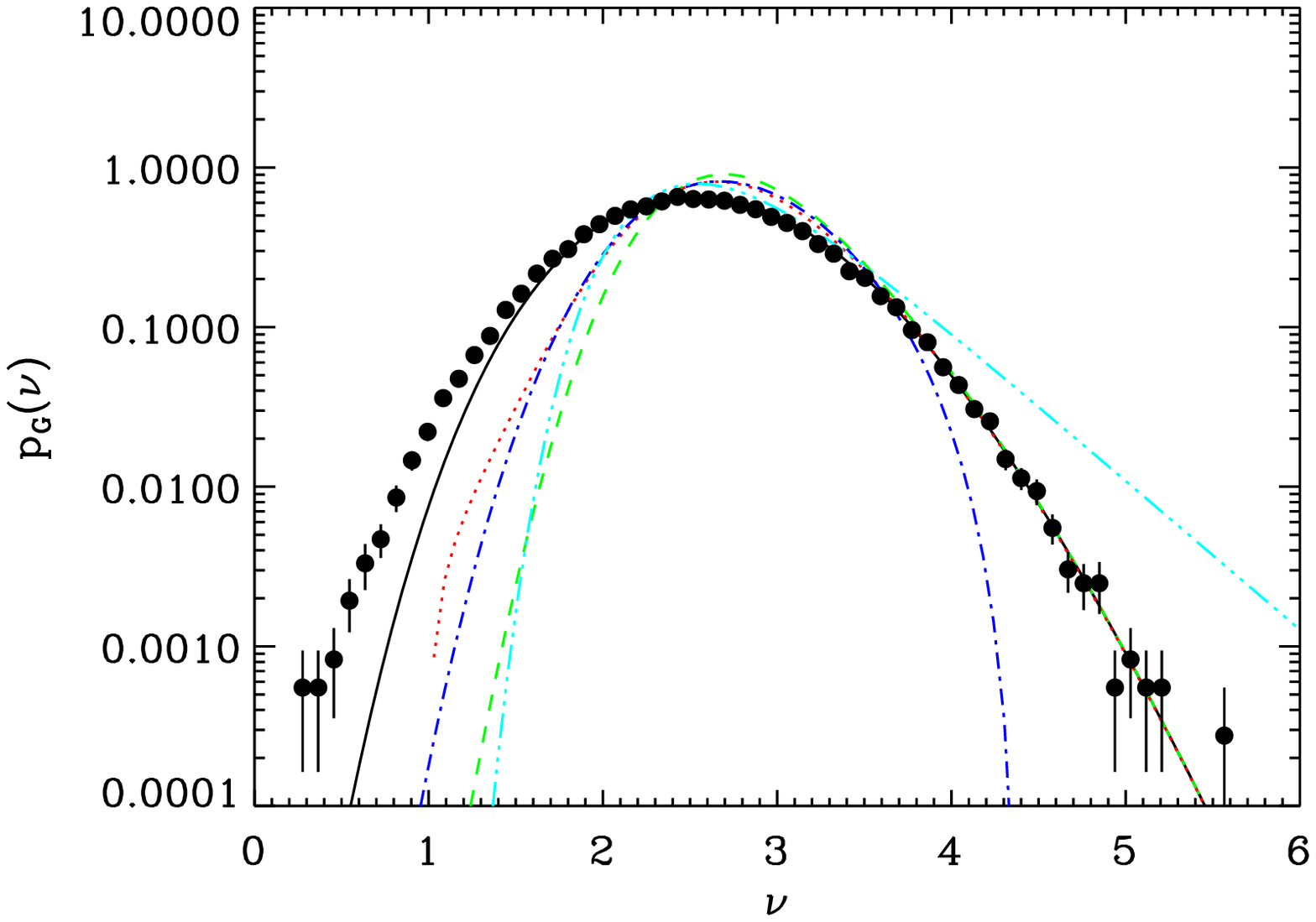}
}}
\centerline{\hbox{
\includegraphics[width=0.5\textwidth]{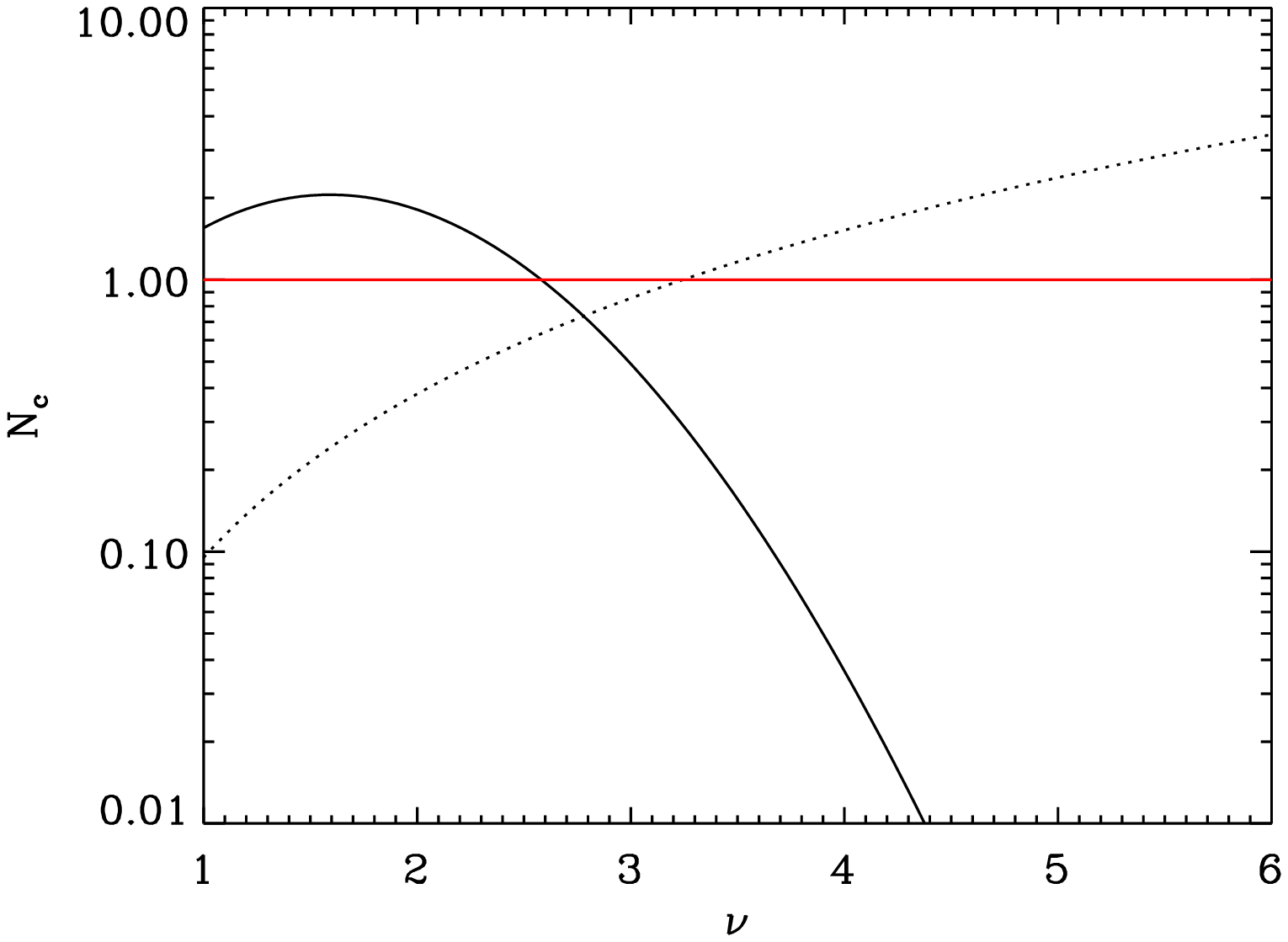}
\includegraphics[width=0.5\textwidth]{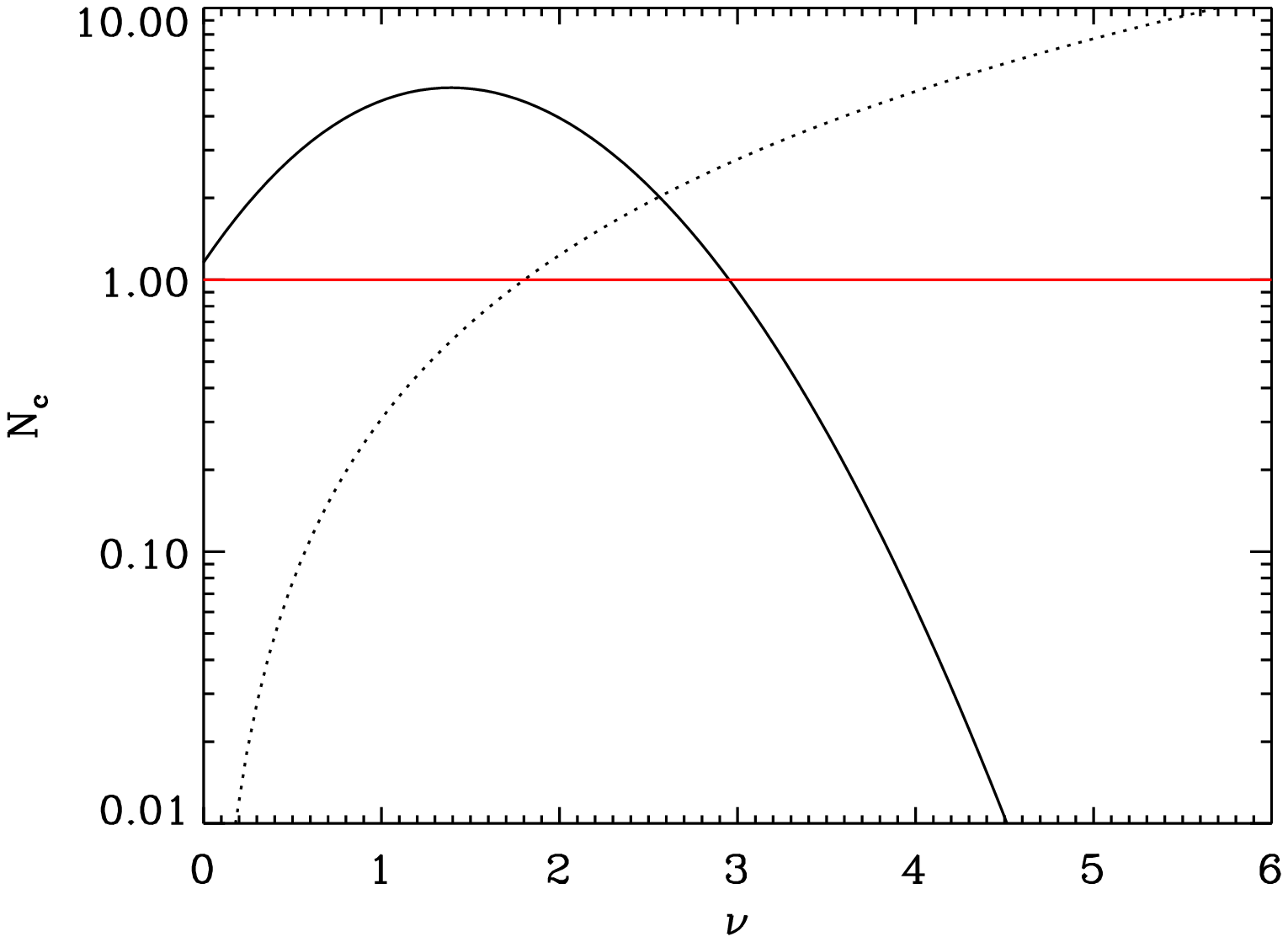}
}}
\caption[]{Same as in Fig.~\ref{fig:pdfs}, but for $n=-1$ and
  $n=-1.5$. \label{fig:pdfs2}}
\end{figure*}

The results are displayed in Figure~\ref{fig:pdfs} for $n=0$, $-0.5$
and in Figure~\ref{fig:pdfs2} for $n=-1.0$ and $-1.5$. Agreement
between the measurements and theory is spectacular--with the best
results in the high-$\nu$ region,
as expected. Even the case $n=-1.5$ is well described by theoretical predictions
despite the fact that condition (\ref{eq:validd}) is broken while
$N_{\rm c} \ga 1$. Except for
$n=-1$, the low-$\nu$ tail is slightly off the theory, which
overestimates a bit the measurements for $n > -1$  and
significantly underestimates them for $n=-1.5$. Still, our analytic calculations
are sufficiently accurate to define confidence regions with an accuracy
level of a few percent. As expected, the effect of clustering between peaks
decreases with increasing $n$ and becomes rather small for $n > -1$, given 
the choice of $L/\ell=20$. In that regime, the
number density of peaks is in fact well approximated by the
Euler Characteristics, and the function $p_{\rm G}(\nu)$ is well fitted
by a negative Weibull type form  (eq.~\ref{eq:gumbform}) 
with the parameters derived in \S~\ref{sec:asymptotic},
except in the high $\nu$ tail, as expected. In agreement with
the predictions of \S~\ref{sec:asymptotic}, $|\gamma_{\rm G}|$
decreases with $n$ and the asymptotic regime (\ref{eq:asigo}) is
approached slowly although not reached yet, especially in the tails. 
It was indeed argued in \S~\ref{sec:asymptotic} that convergence to it is rather
slow and requires an increasingly large value of $L/\ell$
as $-n$ becomes larger. Note that the data points or the solid curves can also be
fitted easily by the form (\ref{eq:gumbform}) with appropriate choice of $a$, $b$ and
$\gamma_{\rm G}$ \citep{Mikelsons}, except of course for the high $\nu$
tail. For simplicity we did not show that fit because it is purely
phenomenological and there is no simple analytic expression for $a$
$b$ and $\gamma_{\rm G}$ except in the asymptotic regime studied in
\S~\ref{sec:asymptotic}.

\section{Summary and discussion}
\label{sec:discussion}
We have computed analytically the Gumbel statistics for random
Gaussian fields smoothed with a Gaussian window of size $\ell$.  
The Gumbel statistics, $p_{\rm G}(\nu_{\rm max}) {\rm d}\nu_{\rm max}$,  
represents the probability distribution
of the maximum $\nu_{\rm  max}$ of the field estimated in a patch of
size $L$ thrown at random.
Our important results can be summarized as follows:
\begin{enumerate}
\item[(i)] For $L$ sufficiently large in front of $\ell$, $\nu_{\rm max}$ can be
  approximated by the maximum value of the density estimated at the
  positions of the peaks included  in the patch. As a result, the cumulative Gumbel distribution,
$P_{\rm G}(\nu) =\int_{-\infty}^{\nu} p_{\rm G}(\nu_{\rm max}) {\rm d}
\nu_{\rm max}$, can be seen as the void probability $P_0$  of finding no
peak with density larger than $\nu$ in the patch. 
\item[(ii)] We have made use of the standard counts-in-cells formalism
  \citep{White,Fry,BS,SzapudiSzalay}, to compute this void
probability as a function of the average number of peaks
above the threshold  in the patch, $n(\nu)V$, 
and their correlation functions averaged over the patch, ${\bar \xi}_N^{\rm p}(L)$. 
These quantities were themselves calculated using results of the
literature: \cite{BBKS} and \cite{BE} to estimate $n(\nu)$ and
${\bar \xi}_2^{\rm p}$; \cite{BernardeauSchaeffer} and \cite{PW} to evaluate higher order
correlations through a hierarchy of normalized cumulants given by
$S_N^{\rm p} \equiv {\bar \xi}_N^{\rm p}/({\bar \xi}_2^{\rm p})^{N-1} =
N^{N-2}$.  Rigorously speaking, these calculations are only valid in the large separation limit, ${\bar \xi}_2^{\rm p} \ll
1$ and for $\nu \gg 1$. They also neglect contributions from
higher order derivatives of the correlation function of the density
field, which can in principle be taken into account following
\cite{Sheth} and \cite{Desjacques}.
 \item[(iii)] In the regime $\nu \gg 1$ and in the Poisson limit,
  $N_{\rm c} \equiv n V {\bar \xi}_2^{\rm p} \ll 1$, the quantity $-\ln(P_{\rm G})$
  is simply given by the Euler Characteristic of the excursion \citep{Aldous}. This allows one
to derive tractable analytical expressions for the Gumbel statistics 
(eqs.~\ref{eq:asy3D}, \ref{eq:asy2D}, \ref{eq:UDgen}). We have shown that in this case
$P_{\rm G}(\nu)$ is well fitted by a negative Weibull type distribution (\ref{eq:gumbform}) (with
$\gamma_{\rm G} < 0$), except in the high $\nu$ tail. As expected,
$\gamma_{\rm G} \rightarrow 0$ when $L/\ell \rightarrow \infty$ and one
converges slowly to the Gumbel type distribution,
eq.~(\ref{eq:asigo}), as shown long ago \citep{BR}. 
\item[(iv)] Our analytical calculations were successfully tested  
against numerical experiments of 2D scale-free Gaussian random fields,
in particular in a regime where both $N_{\rm c} \ga 1$ and ${\bar
    \xi}_2^{\rm p} \ga 1$, i.e.~where the validity of the ``exact''
  calculations mentioned in point (ii) remains questionable.   
\end{enumerate} 
Note that our calculations can be easily extended to non-Gaussian
fields, using e.g. the formalism of \cite{Pogo} to
estimate the number density of peaks, $n(\nu)$, and modifications of
the Press \& Schechter formalism \citep{PS} to compute ${\bar
  \xi}_2^{\rm p}$  in the high $\nu$ regime \citep[see for
instance][and references therein]{DS,Valageas}. The hierarchical
relation $S_N^{\rm p} \simeq N^{N-2}$ should still hold if $\nu \gg 1$
\citep{BernardeauSchaeffer}, as extensively discussed in end of
\S~\ref{sec:gene} and in beginning of \S~\ref{sec:sigma}.
In the Poisson limit and for $\nu \gg 1$,
the result obtained in point (iii) above should still hold: $-\ln(P_{\rm G})$
 should be simply given by the Euler Characteristic, which itself can be
easily estimated in the non-Gaussian case \citep{Pogo,matsu}.
In fact, one expects that $P_{\rm G}$  should
still be well fitted by the family of distributions (\ref{eq:gumbform}) but
with a different value of $\gamma_{\rm G}$ \citep{Mikelsons}. However, 
convergence to the asymptotic form (\ref{eq:asigo}) in the limit $L/\ell
\rightarrow \infty$ remains to be proven. 
In the intermediary regime probed by the Euler  Characteristic, the Gumbel
statistics provides an interesting test of non-Gaussianity, as shown 
experimentally by \cite{Mikelsons} on simulated temperature
maps of the CMB.
In a companion paper \citep{Davis}, we have studied
applications of the Gumbel statistics to clusters of galaxies, where
the quantity of interest is the probability distribution
function of the mass of the most massive cluster in the patch
\citep[see also][]{Holz,Cayon}. 
We plan to apply 2D Gumbel statistics to the analysis of CMB data,
including non-Gaussian corrections,  in the near future.

\section*{acknowledgments}
The authors thank F. Bernardeau for very useful discussions.
OD acknowledges the support of an STFC studentship and  
JD' s research is supported by Adrian Beecroft, the Oxford Martin School and STFC.

\label{lastpage}
\end{document}